\documentclass[preprint]{vgtc}               

\graphicspath{{figures/}} 

\PassOptionsToPackage{svgnames}{xcolor}

\ifdefined\ShowChanges
  \usepackage[markup]{changes}
\else
  \usepackage[final]{changes}
\fi

\usepackage{times}                     

\usepackage{tabu}                      
\usepackage{booktabs}                  
\usepackage{lipsum}                    
\usepackage{mwe}                       
\usepackage{circledsteps}

\usepackage{mathptmx}                  
\usepackage{amsmath}
\usepackage{amssymb}

\newcommand{\FlatCircledText}[1]{\smash{\CircledText{#1}}}
\newcommand{\subfigref}[2]{\hyperref[#1]{Fig.~\ref*{#1}#2}}

\vgtcinsertpkg

\preprinttext{\copyright 2025 IEEE. To appear in the proceedings of IEEE Visualization conference.}

\title{Anchoring and Alignment: Data Factors in Part-to-Whole Visualization}

\author{Connor Bailey* \and Michael Gleicher\thanks{Emails: \texttt{[cbbcbail|gleicher]@cs.wisc.edu}}}

\affiliation{\scriptsize University of Wisconsin-Madison}

\teaser{
  \centering
  \includegraphics[width=\linewidth]{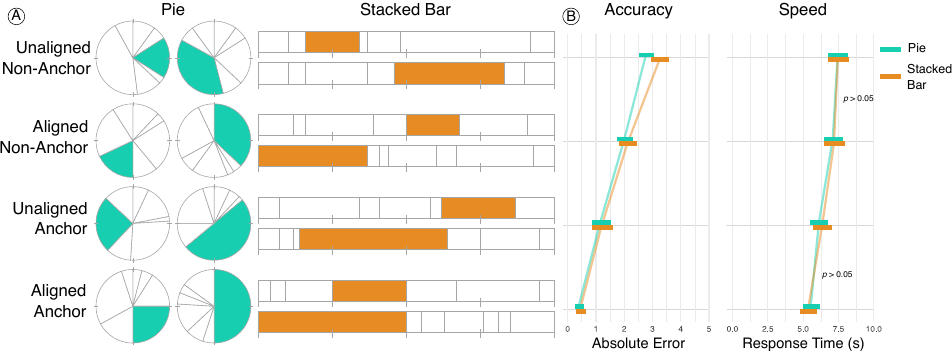}
  \caption{
    \FlatCircledText{A} A pair of example stimuli from each combination of pie and stacked bar charts, aligned (\textit{segment starts/ends at a marker}) and unaligned, and anchor (\textit{segment has a recognizable size}) and non-anchor values.
    \FlatCircledText{B} Predicted mean absolute error and response time to 95\% CI. 
    Anchor values and alignment positions show significant performance improvements in estimating values for both chart types. 
    Results between conditions are significant ($p < 0.05$) except where noted.
  }
  \label{fig:teaser}
}

\abstract{
  We explore the effects of data and design considerations through the example case of part-to-whole data relationships.
  Standard part-to-whole representations like pie charts and stacked bar charts make the relationships of parts to the whole explicit.
  Value estimation in these charts benefits from two perceptual mechanisms: \textit{anchoring}, where the value is close to a reference value with an easily recognized shape, and \textit{alignment} where the beginning or end of the shape is aligned with a marker.
  In an online study, we explore how data and design factors such as value, position, and encoding together impact these effects in making estimations in part-to-whole charts.
  The results show how salient values and alignment to positions on a scale affect task performance.
  This demonstrates the need for informed visualization design based around how data properties and design factors affect perceptual mechanisms.
}


\begin{document}

\maketitle

\section{Introduction} 
The properties of data, and how they relate to data visualization design choices, can have a significant influence on how viewers perform low level tasks. 
However, while there has been a focus on how encoding affects task performance, there has been less attention on how the data factors themselves affect performance. 
In this paper, we examine the task of estimating parts in part-to-whole data relationships as a case study. 
Standard charts such as pie charts and stacked bar charts are commonly used and make the relationships between parts to the whole explicit.
Two perceptual mechanisms affect estimation in these chart types: \textbf{anchoring}, where the value is similar to a reference value with an easily recognized shape, and \textbf{alignment} where the beginning or end of the shape is aligned with a marker allowing for easier judgement.

We explore how data and design factors such as value, position, and chart type together impact these perceptual mechanisms when making estimations with part-to-whole charts, and how this affects task performance.
When estimating the part-to-whole relationship, the viewer compares the value of a part to the nearest anchor value. 
For pie charts, there are natural anchors at 25\% and 50\%; stacked bars have similar anchors from scale markings. 
Additionally, the alignment of a value to a scale marker provides a baseline that should improve the viewer's ability to use the anchor as a reference. 
We therefore hypothesized that anchor values and alignment positions would result in better task performance than non-anchor and unaligned values.
We expected this would also affect values near anchors and positions near alignment, such that values near anchors and positions near alignment would have better task performance than those further away.

In an online study, we examine how salient values, and alignment to positions on a scale affect response time and accuracy in estimating parts in part-to-whole charts. 
We compared pie charts and stacked bar charts as they are standard visualizations that explicitly show part-to-whole relationships.
Additionally, these charts have been used in prior studies, facilitating comparison.
We used a within-subjects, stratified randomized experimental design to measure performance across the large space of conditions.
We fit generalized linear mixed-effects models (GLMMs) to the results to analyze the effects of alignment, anchoring, rounding, chart type, and their interactions on task performance.
We observe that the relationship between the data value being read and the anchor values, and the relationship between the position on the chart and the alignment positions, have a significant impact on task performance.
We find similar patterns across chart type.

Our main contribution is to demonstrate the effects of data and design factors on perceptual mechanisms, and the resulting effects on task performance.
Our work has three types of ramifications. 
First, in terms of visualization design, it provides evidence to support a nuanced approach to making design decisions based on data properties.
Second, it shows the need to consider data and design factors and perceptual mechanisms in designing visualization experiments. 
Third, it contributes to our understanding of the mechanisms of how basic charts are perceived by viewers, showing how understanding of anchoring, and alignment relate to the large variations across different conditions within chart types. 
While we examine standard charts for the low-level task of part-to-whole estimation, we expect that the concepts and methodology generalize to other tasks and visualization design considerations.

\section{Related Work}
The empirical assessment of the connection between encoding and task performance has a long history in visualization research, with seminal work by Cleveland and McGill \cite{cleveland1985GraphicalPerceptionGraphical} and Mackinlay \cite{mackinlay1986AutomatingDesignGraphical}.
Following studies have continued to build on this tradition \cite{quadri2022SurveyPerceptionBasedVisualization,zeng2023ReviewCollationGraphical}. 
Heer and Bostock \cite{heer2010CrowdsourcingGraphicalPerception} 
showed that such studies can be run online, allowing larger experiements with more nuanced results.
Kim and Heer \cite{kim2018AssessingEffectsTask} showed that the 
effects of encoding vary based on task and data.
Our work continues in this direction by focusing on the effects of data and design factors on task performance.

\subsection{Part-to-Whole Studies}
\label{sec:partWhole}

Studies examining performance on charts showing a part-to-whole relationship have been done since at least Eells \cite{eells1926RelativeMeritsCircles} in 1926 who found judgements of pie charts to be more accurate than stacked bar charts. 
Following studies conflicted \cite{vonhuhn1927FurtherStudiesGraphic}, evaluated the proportional judgement task \cite{croxton1927BarChartsCircle, cleveland1985GraphicalPerceptionGraphical, mccoleman2021RethinkingRanksVisual}, and compared these charts with other variations \cite{peterson1954HowAccuratelyAre}. 
In their 1984 study, Cleveland and McGill \cite{cleveland1984GraphicalPerceptionTheory} differentiate between length and position encoding which may explain some variation in prior work. 

Heer and Bostock \cite{heer2010CrowdsourcingGraphicalPerception} replicate the findings of Cleveland and McGill and include missing comparisons from the previous study. 
They show that pie charts outperform length encoding, but not position encoding.
Hollands and Spence \cite{hollands1998JudgingProportionGraphs} compared pie charts with stacked bar charts, bar charts with no scale, and bar charts with a 100\% reference for estimating proportions. 
These studies show that there are complex effects related to value and position in these visualizations which we further explore in our study. 

Kosara examined how people use pie charts to make judgements about parts using variations of charts to make comparisons \cite{kosara2016JudgmentErrorPie, kosara2019CircularParttoWholeCharts}. 
Kosara and Skau \cite{kosara2019EvidenceAreaPrimary} also examined the use of pie charts in a study on the effects of angle, arc length, and area on judgement.
Their findings show nuance in the ways people interpret pie charts, providing evidence that it is not a simple reading of angles, but that area and arc length have significant effects \cite{skau2016ArcsAnglesAreas}. 
We seek to understand more about how part-to-whole charts are read by further exploring the effects of value and alignment on task performance.

\subsection{Mechanisms and Biases}
\label{sec:mechanisms-biases}

\indent\textbf{Anchoring}
generally refers to the bias effect caused by some initial value on the judgement of a decision-making actor \cite{tversky1974JudgmentUncertaintyHeuristics, cho2017AnchoringEffectDecisionMaking, furnham2011LiteratureReviewAnchoring, valdez2018PrimingAnchoringEffects}. 
Simkin and Hastie \cite{simkin1987InformationProcessingAnalysisGraph} apply the concept to chart reading: where the explicit scales or implicit cues (e.g., the quartile cues of a pie segment) act as anchors in the estimation of a part. 
Redmond \cite{redmond2019VisualCuesEstimation} explores the effects of various visual cues on anchoring in a set of studies comparing the performance of pie charts and horizontal stacked bar charts with two segments. 

\textbf{Alignment} refers to a segment having one of its edges at a scale point or axis. 
It connects segments to the scale (even if the scale is just the entirety of the bar). 
In two-part charts, parts can always be aligned while in multi-part charts, not all parts can be aligned. 
Even if it is desirable to align the segment of interest, it may not be possible (e.g., if it isn't known ahead of time). 
Redmond \cite{redmond2019VisualCuesEstimation} also finds that scales affect performance in bar charts (by providing alignment points), but have less effect on pie charts. 
Our work further explores anchoring and alignment in part-to-whole charts by considering them together with chart type and rounding effects.

\textbf{Rounding} has been shown to have significant effects on decision-making \cite{allen2017ReferenceDependentPreferencesEvidence}. 
Converse and Dennis \cite{converse2018RoleProminentNumbers} show that people often choose from a small set of possible answers rather than trying to compute their best precise estimate. 
However, the visualization literature offers little exploration of rounding. 
For example, Davis et al. \cite{davis2024RisksRankingRevisiting} observe ranking effects, but dismiss their importance. 
We found strong evidence, of rounding in the responses, and therefore include a rounding factor as a covariate in our model.

\section{Methodology}
In order to investigate the effects of data and design factors on task performance in part-to-whole visualizations, we explored the effects of alignment, anchoring, and chart type on estimation accuracy and response time.
Our within-subjects design shows participants 96 stimuli generated by stratified random sampling.
The study was preregistered following pilot studies and prior to data collection.\footnote{\url{https://osf.io/e36au}} More information about the methodology, including the stimuli generation code, data, examples, and results of the exclusion criteria is provided in the supplementary materials.\footnote{\url{https://github.com/uwgraphics/PartToWhole}}
The studies were run under an exempted protocol approved by the University of Wisconsin - Madison IRB.

\subsection{Stimuli}
\label{sec:stimuli}
Each stimulus is a pie chart or stacked bar chart of 7 integer values.
One target value (between 15\% and 55\%) was selected to be estimated by the participant.
The target values were generated by stratified random sampling.
The values of the other 6 parts were subsequently drawn from a beta distribution $\text{Beta}(\alpha=2, \beta=10)$ to avoid having many small or large values.
We pre-generated the stimuli, creating a unique set for each participant by drawing a stratified random subset of 96 with equal numbers of each chart type.

Our conditions define \emph{anchor} stimuli as when the target has an exact anchor value (25\% or 50\%), \emph{near-anchor} stimuli as within 4\% of an anchor value, and \emph{far-anchor} as more than 4\%.
Similarly, \emph{alignment} stimuli are when the target segment starts or ends exactly at a marker, \emph{near-alignment} stimuli as within 4\% of a marker, and \emph{far-alignment} as more than 4\% from a marker.
These served as strata of our sampling routine to ensure sufficient sample sizes of each condition.
We sampled 24 \emph{aligned}, 36 \emph{near-aligned}, and 36 \emph{far-aligned} target values for each participant.
Similarly, we sampled 12 \emph{anchor} values, 36 \emph{near-anchor} values, and 48 \emph{far-anchor} target values for each participant.
Examples are shown in \subfigref{fig:teaser}{A}.

\subsection{Procedure}
\label{sec:procedure}

Participants were directed to the consent form of the IRB protocol which described the research, participation, risks, benefits, compensation, confidentiality and contact information. 
Consenting participants were checked for having JavaScript and cookies enabled, necessary to complete the study. 
They manually verified that they were using a standard browser, a large enough screen, and that they resized the browser window to fill the screen.
In order to reduce the differences in the size of the stimuli due to differences in screen sizes, participants resized a rectangle to the size of a standard credit card or driver's license on their screen and the stimuli were proportionally resized accordingly.

Participant training comprised 8 example questions, split evenly by chart type. 
Training provided task instructions, a description of the study process, and feedback on each response. 
This allowed participants to learn the task and the input method, and verify they were doing it correctly.
A simple example scenario was used to ground the task.
The data was described as the sales of smartphones by 7 smartphone companies.
The task was to estimate the percentage of the total for a highlighted company.
The companies are represented by letter labels. 
The target value was highlighted in green, and the remaining parts appeared white.

Each question begins with a screen containing a fixation marker (`+') in the center of the screen for a period of 500 milliseconds. 
The visualization was generated, but hidden, until the 500 milliseconds passed and the question text, input box, and chart appear. 
The current question number, out of the total number of questions, was indicated at the top of the page as a reference to the participant.
Questions were given in sets of 12 that automatically advanced upon submission. 
Every 12 questions the participant was able to pause before pressing enter to advance to the next set of questions.

\textbf{Participant Recruitment:}
We recruited participants through the \href{https://www.prolific.com/}{Prolific} platform. 
Participants spent a median time of less than 20 minutes total to complete the study and were compensated \$4.00 for a rate of ~\$12.00 per hour. 
The study included 60 participants.
The sample size was selected by conducting an \textit{a priori} power analysis by bootstrap sampling a pilot sample of ten participants to obtain 0.9 power to detect a significant effect at the standard $p<\alpha= 0.05$ level. 
We rounded up to an even number of 60 participants to ensure sufficient sample sizes for post-hoc analysis. 
Participants were prescreened through Prolific for fluent English speakers, with no colorblindness, between the ages of 18 and 65, using a desktop computer, and not having participated in any of the previous studies or pilot studies we conducted as part of this work. 

\textbf{Exclusion Criteria:}
We preregistered exclusion criteria to limit the effects of outlier responses and response times and inattentive participants distorting the analysis of our results. 
Participants were excluded for having a mean absolute error outside of two median absolute deviations from the median absolute error. 
Response times were clipped to two standard deviations from the median response time.
Response times over the exclusion threshold were clipped to the threshold instead of dropped in order to maintain balance required for statistical tests. Participants with mean absolute error over the threshold were excluded and back-filled by a replacement participant. 
In our study, 22 participants were excluded and replaced based on the preregistered criteria.

\textbf{Measures:}
We measured \textit{absolute error} as the absolute value of the difference between the participant's response and the correct answer.
Absolute error is in discrete integer values, and non-negative, with a maximum possible error of 85\% (100\% - 15\%).
We also measured \textit{response time} as a secondary measure.
Response time was measured from the moment the participant saw a question to the moment they submitted their response.
Both absolute error and response time are non-normally distributed, and positively skewed. 

\textbf{Hypotheses:}
The perceptual mechanism of anchoring in part-to-whole charts suggested our hypotheses. 
Pie charts have  natural anchors at 25\% and 50\%, and we predicted this would lead to better performance than with stacked bar charts.
\begin{align}
    &\mathsf{H}^\text{error}_\text{chart} : \left|\text{error}\right|_\text{pie} < \left|\text{error}\right|_\text{stacked-bar} \notag
\end{align}
We predicted anchoring and alignment would have a significant effect, where anchor values and aligned positions would result in better performance than non-anchor values and unaligned positions.
\begin{align}
    &\mathsf{H}^\text{error}_\text{anchor} : \left|\text{error}\right|_\text{anchor} < \left|\text{error}\right|_\text{non-anchor} \notag \\
    &\mathsf{H}^\text{error}_\text{alignment} : \left|\text{error}\right|_\text{aligned} < \left|\text{error}\right|_\text{unaligned} \notag
\end{align}
We also hypothesized task performance would decrease with increasing distance from anchor values and aligned positions.
\begin{align}
    &\mathsf{H}^\text{error}_\text{anchor distance} : |\text{error}|_\text{anchor} < |\text{error}|_\text{near-anchor} < |\text{error}|_\text{far-anchor} \notag \\
    &\mathsf{H}^\text{error}_\text{alignment distance} : |\text{error}|_\text{align} < |\text{error}|_\text{near-align} < |\text{error}|_\text{far-align} \notag
\end{align}
We preregistered our hypotheses for absolute error, but also analyzed equivalent hypotheses for response time.
\begin{figure}[tb]
    \centering
    \includegraphics[width=\columnwidth]{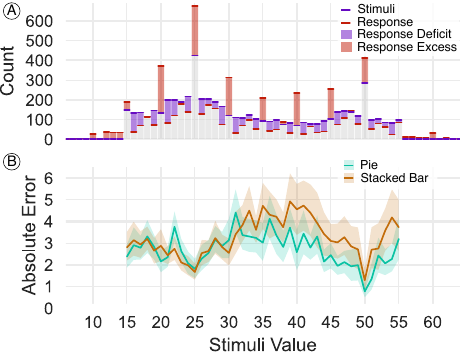}
    \caption{\FlatCircledText{A} Frequency of responses at each stimuli compared to frequency of stimuli values show a significant effect of rounding (overlap in gray). \FlatCircledText{B}Absolute error across the full range of values reveals the anchor effect and the distance to anchor effect near the 25\% and 50\% anchor values.}
    \label{fig:responses}
\end{figure}

\begin{figure*}[t]
    \centering
    \includegraphics[width=\textwidth]{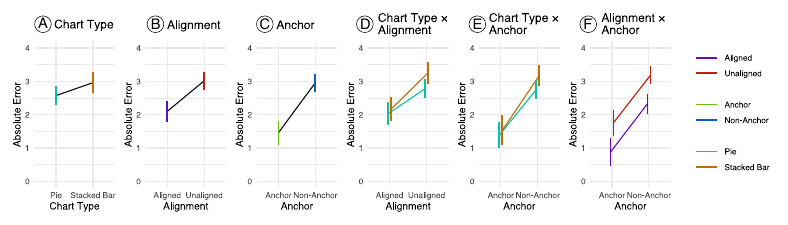}
    \caption{Main effects of \FlatCircledText{A} chart type, \FlatCircledText{B} alignment, and \FlatCircledText{C} anchor value on absolute error from model predictions. Interaction effects of \FlatCircledText{D} chart type and alignment, \FlatCircledText{E} chart type and anchor value and \FlatCircledText{F} alignment and anchor value on absolute error as predicted by the model.}
    \label{fig:results}
\end{figure*}

\section{Results}
We modeled absolute error and response time with generalized linear mixed-effects models (GLMMs) in order to determine significant effects and test our hypotheses.
The models include random effects per participant to account for individual differences in responses.
We used a negative binomial distribution with a log link to model absolute error, as the absolute error is discrete, non-negative, and non-normal. 
We used a Gamma distribution with a log link for response time, as response time was continuous, positive, and non-normal.
Random intercepts and slopes for anchor, alignment, and chart type were included and allowed to correlate within participants. 
Model comparison by Akaike Information Criterion (AIC) indicated substantially better fit for the correlated random effects structure, and the distributions selected.
Details of the model results, fitting process, and summary tables are available in the supplementary materials, here we highlight the key findings. 

To address the confound of rounding, our pre-registered analysis plan involved evaluating the distance to the nearest 5 or 10 as covariates.
AIC model comparison tests showed that including the distance to the nearest 10 significantly improved the model fit over a model with no rounding ($\smash{\Delta AIC}=4.6$) or a model with rounding to the nearest 5 ($\smash{\Delta AIC}=7.35$)
Rounding by 10s was a significant predictor of absolute error in the model ($\smash{\hat{\beta}} = 0.02$, $p<0.05$) and was therefore included.
Visual inspection of the data confirms the presence of rounding (see \subfigref{fig:responses}{A}).

We found a significant difference in absolute error for anchor ($\smash{\hat{\beta}} = 0.66$, $p< 10^{-16}$) and alignment ($\smash{\hat{\beta}} = 0.36$, $p< 10^{-16}$) and a significant difference in response time for anchor ($\smash{\hat{\beta}} = 0.11$, $p<10^{-16}$) and alignment ($\smash{\hat{\beta}} = 0.05$, $p<10^{-16}$).
However, we did not find a significant difference in absolute error or response time for chart type ($p>0.05$).
These results support $\smash{\mathsf{H}_{\text{alignment}}}$ and $\smash{\mathsf{H}_\text{anchor}}$, that aligned positions and anchor values support better task performance in estimating values in a part-to-whole visualization.
We did not find support for $\smash{\mathsf{H}_\text{chart}}$, and differences in task performance between chart types are small. The anchoring effect, and its relation to alignment positions, is similar for viewers of both chart types (see \subfigref{fig:teaser}{B} and \cref{fig:results}).

We found a significant interaction between anchor and alignment for absolute error ($\smash{\hat{\beta}} = -0.17$, $p< 10^{-8}$) and response time ($\smash{\hat{\beta}} = -0.02$, $p<0.01$).
This supports the hypothesis that the effects of anchor values and alignment positions are not independent, i.e. that anchor values are easier to read when aligned.
We did not find a significant interaction between chart type and anchor or chart type and alignment for either absolute error or response time ($p>0.05$).
This suggests that the effects of anchor values and alignment positions are similar for both pie charts and stacked bar charts, and that the slight differences in task performance between the two chart types are due to other factors (see \cref{fig:results}).

We also modeled \emph{anchor}, \emph{near-anchor} and \emph{far-anchor} as ordinal levels in an anchor distance factor and \emph{alignment}, \emph{near-alignment} and \emph{far-alignment} as ordinal levels in an alignment distance factor in a separate model.
We found that both distance to anchor values ($\smash{\hat{\beta}} = 0.62$, $p<10^{-16}$) and distance from alignment position ($\smash{\hat{\beta}} = 0.48$, $p<10^{-16}$) are significant predictors of estimation task performance.
We also found a significant interaction between these two factors ($\hat{\beta} = -0.15$, $p<10^{-8}$), but no significant interactions with chart type ($p>0.05$).
The anchor distance effect can be seen in the responses in \subfigref{fig:responses}{B}.
These findings support $\smash{\mathsf{H}^\text{error}_\text{anchor distance}}$ and $\smash{\mathsf{H}^\text{error}_\text{alignment distance}}$, and suggest that alignment and anchoring are continuous perceptual mechanisms that affect task performance.

\section{Discussion}
The results of the study show the impact of data and design factors on task performance in part-to-whole visualizations.
We found significant performance improvements in estimating \emph{anchor} values and \emph{near-anchor} values, supporting the model where anchoring is a key mechanism of reading the charts.
Our results also support the conventional wisdom recommending \emph{alignment} when possible in chart design, although the effect is small.
We did not find that chart type had a significant effect on performance and so cannot provide a recommendation of chart type based on performance.
These results suggest that data values should be considered in design, in addition to other commonly considered properties like data types, measurement scales, and semantic types.

The task of estimating part-to-whole relationships invokes anchoring and therefore the data and design factors that affect anchoring are important to consider.
While designers typically can't choose their data, our results show the importance of awareness of the effects data will have on performance.
For example, designers can expect that part-to-whole charts with \emph{anchors} or \emph{near-anchor} values will be more quickly and accurately estimated than charts with \emph{non-anchor} or \emph{far-anchor} values.
Data-dependent chart design may prove challenging for the designer, but could be supported by automatic chart generation and recommendation tools.
Our results suggest that exact alignment is not necessary to see performance improvements, and so maximizing the \emph{near-alignment} cases in chart design could prove beneficial.

In empirical study of visualization task performance, it can be important to account for data dependent factors.
Our work shows that a careful consideration of the mechanisms of how viewers perform tasks can reveal how data values relate to those mechanisms, which can have a significant effect on  performance.
Not only do we find that these factors have significant, large effects, and interactions in the study, but that they coincide with several perceptual mechanisms which also significantly affect task performance. 
We found that the inclusion of each of these factors significantly improved the fit of our model to the data.

\subsection{Limitations}
\label{sec:limitations}

Our work focused on the specific task  of part-to-whole estimation with specific chart types (pie and stacked bar charts). 
We speculate that this basic task is a step in analyzing more complex ones, such as remembering multiple values \cite{mccoleman2021RethinkingRanksVisual} or determining ratios \cite{cleveland1985GraphicalPerceptionGraphical}. 
Similarly, we suspect that other chart types have similar data dependencies, possibly caused by similar mechanisms. 
We view our work as a step towards generalizable knowledge by identifying that such data dependent factors exist, in at least one common case.
A more comprehensive understanding of data dependent factors could have implications in data collection, data selection, and chart design.

Our work does not consider the sources of anchors. 
For pie charts, we did not assess the importance of the implicit (shape-based) and explicit (scale-mark) anchors. 
For stacked bars, we did not explore how the anchors might be created by design elements such as the scale marks.
We have observed large variance between viewers across the main effects in the study.
Our work does not connect these variations to the design and data dependencies.
In future work, we hope to find viewer-dependent factors, such as a propensity for rounding, that hold across situations. 

\subsection{Conclusions}
\label{sec:conclusions}

This paper provides an example of measuring the effects of data and design factors on perceptual mechanisms, through measuring task performance.
The study examined common part-to-whole charts, showing the impact of factors (alignment, anchoring, and rounding) beyond chart type.
Data value differences (e.g., distance to anchor), design choices (e.g. alignment), and user behavior (e.g. rounding) all impact performance. 
Anchoring appears to be a key mechanism in perception of these charts, and partially explains their data-dependent performance.
More generally, we believe our approach of trying to understand data-dependent variance can lead to better understanding and guidelines for design of visualizations.

\acknowledgments{
This work was supported in part by NSF award 2007436.}

\bibliographystyle{abbrv-doi-hyperref-narrow}
\bibliography{25-VisShort-PartWhole}

\end{document}